\documentclass[12pt]{article}
\usepackage{amsfonts} 
\usepackage{a4}
\usepackage{epsfig}
\usepackage{latexsym}
\begin{document}

\title {The ground state energy of a massive scalar field in the background of a semi-transparent spherical shell}
\author{Marco Scandurra \thanks{e-mail: scandurr@itp.uni-leipzig.de} \\
  Universit\"at Leipzig, Fakult\"at f\"ur Physik und Geowissenschaften \\
  Institut f\"ur Theoretische Physik\\
  Augustusplatz 10/11, 04109 Leipzig, Germany}
\maketitle

\begin{abstract}
We calculate  the zero point energy of a massive scalar field in the background of an infinitely thin spherical shell given by a potential of the delta-function type. We use the zeta functional regularization and express the regularized ground state energy in terms of the Jost function of the related scattering problem. Then we find the corresponding heat kernel coefficients and perform the renormalization, imposing the normalization condition that the ground state energy vanishes when the mass of the quantum field becomes large. Finally the ground state energy is calculated numerically. Corresponding plots are given for different values of the strength of the background potential, for both attractive and repulsive potentials. 
\end{abstract} 
PACS class 04.62 +v

\section{Introduction}
In the recent years much interest has turned to the study of the Casimir energy for spherical configurations. The issue concerns  in particular the attempt to explain by means of QFT the puzzling phenomenon of sonoluminescence \cite{1}, that is the emission of short intense pulses of light by collapsing bubbles of air in water. Up to now the hypothesis that the Casimir energy could play a role in the photon emission has not been supported enough by a satisfactory QFT-model for the dielectric {\itshape ball}.  To calculate the Casimir energy for such a ball could be of great importance to compare, in the statical approximation, the order of magnitude of this energy with the effectively observed light pulses . Many authors have already studied  a perfectly conducting spherical shell in the vacuum of the electromagnetic field, we cite here only paper \cite{2},  other authors have investigated the conducting shell for massive fields \cite{3}. 
In this paper we propose the more realistic model of a semi-trasparent shell. In calculating the ground state energy we use zeta regularisation techniques and Jost function approach. The heat kernel coefficients for a penetrable
sphere have alreay been calculated in a very recent work \cite{4}. We  carry out the complete calculations in order to find the exact numerical dependence of the ground state energy on the radius of the sphere.
\section{The model}
We want to study the ground state energy (GSE) of the scalar field
$\varphi(t,\vec{x})$ (to be quantized) in the background of a potential $V(r)$ .
We will consider the following field equation
\begin{equation}
(\Box +m^2+V(r))\varphi(x)=0,
\end{equation}
where $m$ is the mass of the field. 
The spherical shell is a geometrical object with radius $R$ and a surface $S$, to whom it can be associated a classical energy in terms of classical parameters. 
The total energy of the system reads
\begin{eqnarray}
E & = & E_{class}+E_{quant}\nonumber \\
        & = & \left(pV + \sigma S + FR +k +\frac hR\right) +\left({1\over 2}\sum _{n} \omega _{n}\right),
\end{eqnarray} 
where $V$ is the volume of the sphere, $p$ is the pressure, $\sigma$ is the surface tension and $F$, $k$ and $h$ are other parameters. The classical energy is expressed in a general dimensionally suitable form  which depends on powers of $R$, this definition is useful (as we will discuss later) for its renormalization. The quantum contribution in (2) is the traditional expression for the vacuum energy of a scalar field whose energy eigenvalues are  $\omega_n$ . To render the  eigenvalues of the energy  discrete we take a finite quantization volume with radius  $L \gg 1$.\\
The classical shell is static and spherically symmetric. It is described by a potential
\begin{equation}
V(r)= \frac{\alpha}{R} \delta (r-R),
\end{equation}                 
where $\alpha$ is the magnitude of the potential. The semi transparency of the boundary will be imposed later.
The quantum contribution to the total energy is divergent, for the regularization  we adopt a zeta function technique. We define a regularized  ground state energy 
\begin{equation}
E_{\varphi}={1\over 2}\sum _{(n)} (\lambda _{(n)}^2 + m^2)^{1/2-s} \mu ^{2 s} . 
\end{equation}
where $\mu$ is an arbitrary mass parameter, $s$ is the regularization parameter which we will put to zero after renormalization and
$\lambda_{(n)}$ are the eigenvalues of the wave equation
\begin{equation}
[-\Delta + V(x)] \phi _{(n)}(x)=\lambda _{(n)}^2  \phi _{(n)}(x) .
\end{equation}
Now we introduce a zeta function. The zeta function of the wave operator with potential $V(r)$ as defined in (5) is
\begin{equation}
\zeta_V(s)=\sum_{(n)}(\lambda _{(n)}^2 + m^2)^{-s},
\end{equation}
then we can express the ground state energy in terms of the zeta function
\begin{equation}
E_{\varphi}=\frac12\zeta_V(s-\frac12)\mu^{2s}.
\end{equation}
Since is valid the equalty
\begin{equation}
x^{-s}=\frac{1}{\Gamma (s)}\int^{\infty}_0 dt\ t^{s-1} e^{-xt},
\end{equation}
we can write eq.(6) in the following form
\begin{equation}
\zeta_V(s)=\sum_{(n)}\frac{1}{\Gamma (s)}\int^{\infty}_0 dt\ t^{s-1} e^{-\lambda_{(n)}t-m^2t},
\end{equation}
or
\begin{equation}
\zeta_V(s)=\frac{1}{\Gamma(s)}\int^\infty _0 dt\ t^{s-1}e^{-m^2t}\sum_{(n)} e^{-\lambda_{(n)}t},
\end{equation}
that is 
\begin{equation}
\zeta_V(s)=\frac{1}{\Gamma (s)}\int^{\infty}_0 dt\ t^{s-1}e^{-m^2t} K(t).
\end{equation}
The function $K(t)$ represents the heat kernel.
Now taking the asymptotic expansion of the heat kernel for $t\rightarrow 0$   
\begin{equation}
K(t)= \sum _{(n)} \exp (- \lambda _{(n)}^2 t)\stackrel{t\rightarrow 0}{\sim} \left(\frac{1}{4 \pi t}\right) ^{3/2}  \sum^{\infty}_{j=0} A_j t^j; \ \ \ \ j=0,\frac12,1...
\end{equation}
and making the substitution $s\rightarrow s-1/2$ in eq.(11), we get an expansion of $E_{\varphi}  $  in which it's easy to isolate all the divergent (pole) terms. This makes possible to define a total divergent contribution to the ground state energy
\begin{eqnarray} 
E_{\varphi}^{div} &  =  & -\frac{m^4}{64 \pi ^4}\left( \frac1                                s + \ln \frac{4\mu ^2 }{m^2} -                                 \frac 12\right) A_0\  -\frac{m^3}{24\pi^{3/2}}A_{1/2} \nonumber \\
                        &     &  +\frac{m^2}{32 \pi ^4}\left( \frac                                1s + \ln \frac{4\mu ^2 }{m^2} - \                                 1\right) A_1 \ + \frac{m}{16 \pi^{3/2}}A_{3/2}  \nonumber \\
                        &     &  -\frac{1}{32 \pi ^2}\left( \frac 1                                s + \ln \frac{4\mu ^2 }{m^2} - \                                 2\right) A_2. 
\end{eqnarray}
The $A_j$ are the heat kernel coefficients. In the definition (13) we have included the fractionary terms $A_{1/2}$ and $A_{3/2}$, which do not contain poles, to satisfy a renormalization condition (see below). Actually only the first three terms seem to contribute to the ultraviolet divergencies. In a smooth background potential their corresponding cofficients are well known:
\begin{eqnarray}
A_0  & = & \int d^3x\ ; \nonumber \\
A_1  & = & -\int d^3x V(x)\ ;\nonumber \\
A_2  & = & \frac12 \int d^3x V^2(x).
\end{eqnarray}
The renormalized zero point energy can be defined as
\begin{equation}
E_{\varphi}^{ren}=E_{\varphi}-E_{\varphi}^{div}.
\end{equation}
To keep the total energy of the system unchanged we add the subtracted object to the classical energy. Then we have also a  definition of a new  classical energy
\begin{equation}
\epsilon_{class}=E_{class}+E_{\varphi}^{div}.
\end{equation}
The transition from $E_{class}$ to $\epsilon_{class}$ consists in the renormalization of the classical parameters contained in eq.(2). Since the heat kernel coefficients are geometrical coefficients depending on the background (and in our case containing powers of $R$), to each classical parameter in the classical energy will correspond dimensionally a term in $E_{\varphi}^{div}$, then we renormalize the parameters as follows
\begin{eqnarray}
p & \rightarrow &  p -\frac{m^4}{64 \pi ^4}\left( \frac1                                s + \ln \frac{4\mu ^2 }{m^2} -                                 \frac 12\right)\ ;\nonumber \\
\sigma & \rightarrow & \sigma -\frac{m^3}{24\pi^{3/2}}\ ; \nonumber \\
F & \rightarrow &  F +\frac{m^2}{32 \pi ^4}\left( \frac                                1s + \ln \frac{4\mu ^2 }{m^2} - \                                 1\right)\ ; \nonumber \\
k & \rightarrow &  k + \frac{m}{16 \pi^{3/2}}\ ; \nonumber \\
h & \rightarrow &  h -\frac{1}{32 \pi ^2}\left( \frac 1                                s + \ln \frac{4\mu ^2 }{m^2} - \                                 2\right). 
\end{eqnarray}
However in our particular case the $A_0$ and    $A_{1/2}$ coefficients, corresponding respectively to $p$ and $\sigma$ , will turn out to  be zero, then only the last three parameters in (17)  will undergo  renormalization. These new parameters are to be inserted in the classical energy in (2). The old classical parameters are infinite quantities (i.e. unphysical) because they don't take into account the vacuum fluctuations.
Before applying this renormalization procedure we must note that the ground state energy proposed in (15) has not yet a unique meaning. For the uniqueness of $ E_{\varphi}^{ren}$ we impose the renormalization condition
\begin{equation}
\lim_{m\rightarrow \infty} E_{\varphi}^{ren} \ =\ 0,
\end{equation}
which physically means that for a field of infinite mass we have no quantum fluctuations. We fulfil such  requirement by subtracting all the contributions in $E_{\varphi}^{div} $ proportional  to positive or zero powers of the mass. That is we  subtract also ''fractionary,, terms up to the term  resulting from the heat kernel coefficient $A_2$. The remaining part, containing negative powers of $m$, will go to zero for $m\rightarrow \infty$.\\  
This is our renormalizaton scheme, now we turn to the calculations. 

\section{Representation of the ground state energy in terms of the Jost function}
We use an approach appeared for the first time in \cite{6}. In the background of a spherical potential we have a radial Schr\"odinger  equation
\begin{equation}
\left( \frac{d^2}{dr^2} -\frac{l(l+1)}{r^2}-V(r)+\lambda^2_{n,l}\right)\phi_{n,l}(r)=0,
\end{equation}
where $l$ is the angular momentum.
In the general scattering theory  with a continous spectrum $p$ we have the ''regular solution,, \cite{7} defined as
\begin{equation}
\phi_{p,l}(r)\stackrel{r\rightarrow 0}{\sim} j_l(pr) ,
\end{equation}
where $j_l(pr) $ is the Riccati Bessel function. The asymptotics of the regular solution is expressed in terms of the Jost function $f_l(p)$ 
\begin{equation}
\phi_{p,l}(r)\stackrel{r\rightarrow \infty}{\sim}  \frac i2\left( f_l(p)\hat{h}^-_l(pr)-  f^{\star}_l(p)\hat{h}^+_l(pr)\right),
\end{equation}
where $\hat{h}^{\pm}_l(pr)$ are the Riccati-Hankel functions. Now we examine the field at the boundary of our quantization volume. The potential has a compact support then, at the boundary, expression (21) becomes an exact equation, which can be considered as an equation for the eigenvalues $p=\lambda_{n,l}$. Now taking for instance Dirichlet boundary conditions: $\phi_{p,l}(L)=0$, we will have 
\begin{equation}
\left( f_l(p)\hat{h}^-_l(pL)-  f^{\star}_l(p)\hat{h}^+_l(pL)\right)=0. 
\end{equation}
We see that eq.(22) is satisfied for $p=\lambda_{l,n}$, then we can write the sum in (4) as a contour integral using Cauchy theorem
\begin{eqnarray}
E_{\varphi} & = & \mu^{2 s}\sum ^\infty _{l=0}(l+ 1/2)\int_{\gamma} \frac{dp}{2\pi i} (p^2+m^2)^{1/2-s} \nonumber \\
            &   & \frac{\partial}{\partial p} \ln \left(f_l(p)\hat{h}^-_l(pL)-  f^{\star}_l(p)\hat{h}^+_l(pL)\right),
\end{eqnarray}
where the contour $\gamma$ encloses all the solutions of eq.(22) on the positive real $p$ axe and also the bound state solutions in the limit $L\rightarrow \infty$, which lie on the imaginary axes . We further simplfy eq.(23) by separating the contour in two pieces $\gamma_1$ and $\gamma_ 2$ and developing the Hankel functions for large $L$. Then it is possible to recognize in the integrand a term $ipL$ which corresponds to the Minkowski Raum contribution. This term can be dropped. Now  we shift the two contours $\gamma_1$ and $\gamma_ 2$ to the imaginary axe and  substitute $p\rightarrow ik$, so that we get
\begin{equation}
E_{\varphi}=-{\cos{\pi s}\over \pi} \mu^{2 s}
\sum ^\infty _{l=0}(l+ 1/2)\int^\infty_m dk [k^2 -m^2]^{1/2 -s} 
\frac{\partial}{\partial k} \ln f_l(ik)
\end{equation}
Since at the end our quantization volume will go to infinity ( $L\rightarrow\infty$ ) this equation will be independent from the boundary condition chosen for the quantization volume. Eq.(24) is a very general and usefull representation of the ground state energy, where all the information about the background potential is contained in  $ f_l(ik)$.\\
In order to perform the substitution $s=0$ and the subtraction proposed in (15) we split eq.(24) 
in to two suitable parts, one of which is divercence free. We obtain this by adding and subtracting the leading uniform asymptotics of the integrand in eq. (24) (for more details on this procedure see for instance \cite{5}). We define
\begin{equation}
E^{ren}_\varphi=E_f + E_{as},
\end{equation}
\begin{equation}
E_f=-{\cos{\pi s}\over \pi} \mu^{2 s}\sum_{l}(l+\frac12)
\int^\infty_m dk [k^2 -m^2]^{1/2 -s} 
\frac{\partial}{\partial k} [\ln f_l(ik)- \ln f^{as}_l(ik)] 
\end{equation}
and
\begin{equation}
E_{as}=-{\cos{\pi s}\over \pi} \mu^{2 s}\sum_{l}(l+\frac12)
\int^\infty_m dk [k^2 -m^2]^{1/2 -s} 
\frac{\partial}{\partial k}  \ln f^{as}_l(ik)-E^{div}_{\varphi} ,
\end{equation}
where $ f^{as}_l(ik)$ is the asymptotics of the modified Jost function which we will take up the third order in $\nu=l+1/2$. As the reader can check, the quantity $E^{ren}_{\varphi}$ introduced  in (15) remains unalterated.\\

Now we need the  modified Jost function corresponding to  our scattering problem, so we turn to study the potential.
\section{Jost function of the $\delta$-shell}
The initial field equation 
\begin{equation}
(\Box +m^2+V(r))\varphi (r)=0
\end{equation}
valid for $-\infty<r<\infty $
can be divided into two parts: an equation for the free field
\begin{equation}
{\rm at}\ \  r\neq R\  \longrightarrow\  (\Box +m^2)\varphi (r)=0  
\end{equation} 

and an equation for the field on the shell, which includes the required transparency conditions 
\begin{equation}
{\rm at}\ r=R\left\{ \begin{array}{ll}\phi '(R+0) - \phi '(R-0) =\ \frac{\alpha}{R}\phi (R)\\  \phi\rightarrow \ {\rm continuous} \end{array} \right.
\end{equation}

For the delta potential  the regular solution  is
\begin{equation}
\phi_{k,l}(r)= j_l(kR)\Theta (R-r)+\frac i2\left(f_l(k)\hat h^-_l(kR)-f^{\star}_l(k)\hat h^+_l(kR)\right)\Theta (r-R)
\end{equation}
for the field inside and outside the radius $R$. As above $ j_l(kR)$ is the Riccati-Bessel function and $ h^{\pm}_l(kR)$ are the Riccati-Hankel functions.  Combining eq.(30) with eq.(31) we get 
\begin{eqnarray}
j_{l,k}(r) &  = & \frac i2 \left(f_l(k)\hat h^-_l(kR)-f^{\star}_l(k)\hat h^+_l(kR)\right), \nonumber \\  
\frac{\alpha}{R}\ j_{l,k}(r) &  = & k\frac i2 \left(f_l(k)\hat h'^-_l(kR)-f^{\star}_l(k)\hat h'^+_l(kR)\right). 
\end{eqnarray}
We solve for $f_l(k)$, keeping in mind that the Wronskian determinant of $\hat h^{\pm}_l$ is $2i$, then  we find
\begin{equation}
f_l(k)=\frac{1}{2i}\left(-2i (-1) + 2i \frac{\alpha}{kR}j_l(kR)\hat h^+_l(kR)\right)
\end{equation}
or
\begin{equation}
f_l(k)= 1+ \frac{\alpha}{kR}j_l(kR)\hat h^+_l(kR)
\end{equation}
and for the modified Jost function we get 
\begin{equation}
f_\nu (ik)= 1+ \alpha I_\nu(kR) K_\nu(kR),
\end{equation}
which is in terms of the modified Bessel functions $I_\nu$ and $K_\nu$ where $\nu=l+ 1/2 $ . We need also the asymptotics of the Jost function: $f_\nu ^{as}(ik)$, or more exactly the logarithmus of $f_\nu ^{as}(ik)$. The expansion of the product of the two Bessel functions in (35), for $k$ and $\nu$ equally large, is easily obtained with the help of \cite{8}. Then we find the needed asymptotics as a sum of negative powers of $\nu$ with coefficients $X_{j,n}$ depending on $\alpha$. The asymptotics up to the third order reads
\begin{eqnarray}
\ln f_\nu ^{as}(ik) & \equiv & \sum_{j,n}X_{j,n}\frac{t^j}{\nu^n}\nonumber \\
&=& \frac{\alpha}{2}\frac{t}{\nu} - \frac{\alpha^2}{8}\frac{t^2}{\nu^2} +\frac{\alpha}{16}\frac{t^3}{\nu^3}+\frac{ \alpha^3}{24}\frac{t^3}{\nu^3} - \frac{3 \alpha}{8}\frac{t^5}{\nu^3} +  \frac{5 \alpha}{16}\frac{t^7}{\nu^3} + O\left(\frac1\nu\right)^4
\end{eqnarray}
Here as above is $\nu=l+ 1/2 $ and  $t=1/\sqrt{1+\frac{k^2R^2}{\nu^2}}$. Now the ground state energy $ E^{ren}_\varphi =E_f+E_{as}$ can be explicit  calculated.\\
Let us first begin with an analytical simplification of $E_{as}$. We transform  the sum over $l$  in an integral with the help of the formula
\begin{equation}
\sum^{\infty}_{l=0}F(l+\frac12)=\int^\infty _0 d\nu F(\nu) +\int^\infty _0\frac{d\nu}{1+e^{2\pi\nu}}\frac{F(i\nu)-F(-i\nu)}{i},
\end{equation}
valid for a each continuous function $F(\nu)$. In our case is
\[
F(\nu)=\int^\infty_m dk\ \nu\  [k^2 -m^2]^{1/2 -s} 
\frac{\partial}{\partial k}  \ln f^{as}_\nu (ik).
\]
Therefore $E_{as}$ is split into two addends
\begin{equation}
E^{(1)}_{as}=-{\cos{\pi s}\over \pi} \mu^{2 s}\int^\infty_0 d\nu
\ F(\nu)
\end{equation}
and
\begin{equation}
E^{(2)}_{as}=-{\cos{\pi s}\over \pi} \mu^{2 s}\int^\infty_0  \frac{d\nu}{1+e^{2\pi\nu}} \frac{(F(i\nu)-F(-i\nu))}{i}\ .
\end{equation}
For the integrations we use the formula
\begin{equation}
\int^\infty _0 d\nu\int^\infty_m dk\ [k^2 -m]^{1/2 -s} 
\frac{\partial}{\partial k} \frac{t^j}{\nu^n}=\frac{(mR)^{2-n}}2\frac{\Gamma (\frac32-s)\Gamma (1+\frac{j-n}2)\Gamma (s+\frac{n-3}2)}{\Gamma (\frac j2)}. 
\end{equation}
First we calculate  $E^{(1)}_{as}$:
\begin{eqnarray}
E^{(1)}_{as} & = & -{\cos{\pi s}\over \pi} \mu^{2 s} \int^\infty _0 d\nu\ \nu \int^\infty _m dk\ [k^2 -m]^{1/2 -s} 
\frac{\partial}{\partial k} \ln f^{as}_\nu(ik))\nonumber \\
             & = & \left(\frac{m^{1-2s}\mu^{2 s}}{\pi}\right) \sum_{j,n}X_{j,n}\frac{(mR)^{2-n}}2\frac{\Gamma (\frac32-s)\Gamma (1+\frac{j-n}2)\Gamma (s+\frac{n-3}2)}{\Gamma (\frac j2)}\nonumber .
\end{eqnarray}
Here, inserting the coefficients of (36) and expanding up to the first order in $s$ all the terms which depend on the renormalization parameter we get 
\begin{eqnarray}
E^{(1)}_{as} &  =  & \frac{2\alpha^3-\alpha}{96\pi R}\left(\frac1s +\ln \frac{4\mu^2}{m^2} -2\right)\nonumber \\
             &    & -\frac{Rm^2\alpha}{8\pi}\left(\frac1s +\ln \frac{4\mu^2}{m^2} -1\right)\nonumber \\
             &    & +\frac{m\alpha^2}{16},\nonumber
\end{eqnarray}
the terms proportional to $m^2$ and $m^0$ contribute to the divergencies of the energy. They are used to calculate the heat-kernel coefficients in (13) and will disappear after the subtraction of $E_\varphi^{div}$. The term proportional to $m$ corresponds to the $A_{3/2}$ term of the heat kernel expansion; although this term seems to generate no divergency it will be as well subtracted as mentioned in the second section of this paper, because of our initial renormalization condition . \\
Now we calculate $E^{(2)}_{as}$ :
\begin{eqnarray}
E^{(2)}_{as} & = & -{\cos{\pi s}\over \pi} \mu^{2 s}\int^\infty _0\frac{d\nu\ \nu}{1+e^{2\pi\nu}}\left(\int^\infty_m dk\ [k^2 -m]^{1/2 -s}\frac{\partial}{\partial k} \ln f^{as}_\nu(ik)+\ C.C. \right)\nonumber \\
             & = &\left(\frac{m^{1-2s}\mu^{2 s}}{\pi}\right) \Gamma\left(\frac32-s\right) \sum _{j,n}X_{j,n} \frac{\Gamma (s+\frac{j-1}{2})}{\Gamma \left(\frac j2\right)(Rm)^j}\left(\int^R _0\frac{d\nu\ \nu}{1+e^{2\pi\nu}}\frac{\nu^{j-n} 2\cos [\pi(j-n)]}{[1-\frac{\nu^2}{m^2R^2}]^{s+\frac{j-1}{2}}}\right.\nonumber \\
             &   & + \left.\int^\infty _R\frac{d\nu\ \nu}{1+e^{2\pi\nu}}\frac{\nu^{j-n} 2\cos [\pi\left(\frac{n-1}{2}-s\right)]}{[\frac{\nu^2}{m^2R^2}-1]^{s+\frac{j-1}{2}}}\right),
\end{eqnarray}
then 
\begin{eqnarray}
E^{(2)}_{as} &  =  &-\frac{1}{32\pi^2}\left(\frac1s +\ln \frac{4\mu ^2 }{m^2}-2\right) -\frac{\alpha}{2\pi R}\ h_1(Rm)\nonumber \\ 
             &     &  -\frac{\alpha^2}{8R^2m}\  h_2(Rm)\nonumber \\
             &     &  + \frac{2\alpha^3+3\alpha}{48\pi R}\  h_3(Rm)\nonumber \\
             &    &  -\frac{\alpha}{8\pi R}\  h_4(Rm)\nonumber \\ 
             &    &  +\frac{\alpha}{48\pi R}\  h_5(Rm), 
\end{eqnarray}
where $h_n(x)$  are the following functions coming from the $\nu$-integrations in eq.(41)
\begin{eqnarray}
 h_1(x)         &  =  & \int^\infty _0 d\nu \frac{\nu}{1+e^{2\pi\nu}} \ln |1-\frac{v^2}{x^2}|\ ; \nonumber \\ 
 h_2(x)         &  =  & \int^{x}_0 d\nu \frac{\nu}{1+e^{2\pi\nu}} \frac{1}{1-\frac{v^2}{x^2}}\ ; \nonumber \\ 
 h_3(x)         &  =  & \int^\infty_0 d\nu \ln|1-\frac{\nu^2}{x^2}|\left(\frac{\nu^2}{1+e^{2\pi\nu}}\right)'\ ;\nonumber \\
h_4(x)         &  =  & \int^\infty_0 d\nu \ln|1-\frac{\nu^2}{x^2}|\left(\frac1\nu\left(\frac{\nu^2}{1+e^{2\pi\nu}}\right)'\right)' \ ;\nonumber \\
h_5(x)         &  =  & \int^\infty_0 d\nu \ln|1-\frac{\nu^2}{x^2}|\left(\frac1\nu\left(\frac1\nu\left(\frac{\nu^2}{1+e^{2\pi\nu}}\right)'\right)'\right)'\ .
\end{eqnarray}
We see that also $E^{(2)}_{as} $ have a pole of the form $\frac1s$ for $j=n=1$. This pole will contribute to the heat kernel coefficient $A_2$. \\
Now, since $E_f$ contains no poles, we are able to write down the complete heat kernel coefficients $A_j$, up to the order $j\leq 2$: 
\begin{eqnarray}
 & A_0= 0;              & \  A_{1/2}=0;\nonumber \\  
 & A_1=- 4 \pi R\alpha; & \  A_{3/2}=\pi^{3/2}\alpha^2;\ \  A_2=- \frac23\pi\frac{\alpha^3}{R}\ .\nonumber
\end{eqnarray}
This coefficients are the same as in paper \cite{4}.
After performing the subtraction
\[
(E^{(1)}_{as}+E^{(2)}_{as})\  -\ E^{div}
\]
$E^{(1)}_{as}$ cancels completely  and only  $E^{(2)}_{as}$ (whitout its divergent portion) will contribute to the total energy, then we have 

\begin{eqnarray}
E_{as} \mid_{s=0} &  = & -\frac{\alpha}{2\pi R}\ h_1(Rm)\nonumber \\ 
                                 &    & -\frac{\alpha^2}{8R^2m}\  h_2(Rm)\nonumber \\
                                 &    & + \frac{2\alpha^3+3\alpha}{48\pi R}\  h_3(Rm)\nonumber \\
                                 &    & -\frac{\alpha}{8\pi R}\  h_4(Rm)\nonumber \\ 
                                 &    & +\frac{\alpha}{48\pi R}\  h_5(Rm)\ . 
\end{eqnarray}
\section{Asymptotics of $E_{as}$}
As a test for the plots that we're going to do, we  check  analitically the  behaviour of  $E_{as}$ for small and for large values of $R$. To do so we must find the corresponding asymptotics of the $h_n(x)$ functions.\\
For $R\rightarrow 0$ we find:
\begin{eqnarray}
\lim_{R\rightarrow 0} h_1(Rm) & \sim & \ln (Rm)\cdot\frac{1}{48}+ C_1\ ;\nonumber\\
\lim_{R\rightarrow 0} h_2(Rm) & \sim &  R^2m^2+ C_2\ ;\nonumber\\
\lim_{R\rightarrow 0} h_3(Rm) & \sim & -2\ln (Rm)\cdot\left(-\frac{1}{2}\right)+ C_3\ ;\nonumber\\
\lim_{R\rightarrow 0} h_4(Rm) & \sim & -2\ln (Rm)\cdot\left(-1\right)+ C_4\ ;\nonumber\\
\lim_{R\rightarrow 0} h_5(Rm) & \sim  & -2\ln (Rm)\cdot\left(-4\right)+ C_5,
\end{eqnarray}
where the $C_n$'s are numbers resulting from the $\nu$-integrations.
Then we have 
\begin{equation}
\lim_{R\rightarrow 0} E_{as}\sim -\frac{\alpha C_0}{16\pi R}-\frac{\alpha^3}{24\pi R}\left(\ln\frac{1}{Rm}-C_3\right) + O\left( \frac1R\right)^1,
\end{equation}
where $C_0=(-8C_1+C_3-C_4+ C_5/3)\sim 0.224$ and $C_3\sim 1.96$.\\
For $R\rightarrow \infty$ we have
\begin{eqnarray}
\lim_{R\rightarrow \infty} h_1(Rm) & \sim & -\frac{1}{m^2R^2}\frac{7}{1920}\ ;\nonumber\\
\lim_{R\rightarrow \infty} h_2(Rm) & \sim &  \frac{1}{48}\ ;\nonumber\\
\lim_{R\rightarrow \infty} h_3(Rm) & \sim &  \frac{1}{24m^2R^2}\ ;\nonumber\\
\lim_{R\rightarrow \infty} h_4(Rm) & \sim &  -\frac{7}{480m^4R^4}\ ;\nonumber\\
\lim_{R\rightarrow \infty} h_5(Rm) & \sim &  -\frac{16}{m^6R^6}\frac{31}{16128},
\end{eqnarray}
then we find
\begin{equation}
\lim_{R\rightarrow \infty} E_{as}\sim -\frac{1}{384}\frac{\alpha^2}{mR^2}+O\left(\frac1R\right)^2.
\end{equation}

\section{Numerical results} 
To study numerically the behaviour of $E_{as}$ 
we write it in the form
\begin{eqnarray}
E_{as} & =  & \frac{\alpha}{16\pi R} g_1(Rm)\nonumber \\
             &    & +\frac{\alpha^2}{16 R} g_2(Rm)\nonumber \\
             &    & +\frac{\alpha^3}{16\pi R} g_3(Rm),
\end{eqnarray}
where the three functions $g_n(x)$ are given by
\begin{eqnarray}
g_1(x) & = & -8h_1(x) +h_3(x) -2h_4(x) + \frac13 h_5(x)\ ;\nonumber \\
g_2(x) & = & -2\frac{h_2(x)}{x}\ ; \nonumber \\
g_3(x) & = & \frac23 h_3(x).
\end{eqnarray}
As it's clear from (49),  for small values of $\alpha$ \ $E_{as}$ will behave like function  $g_1(x)$. For large values of $\alpha$, 
 $E_{as}$ will behave like  function $g_3(x)$. \\
The plots of the three $g_n(x)$ functions are shown below in Fig. 1

\begin{figure}[ht]\unitlength1cm
\begin{picture}(6,6)
\put(-0.5,-5){\epsfig{file=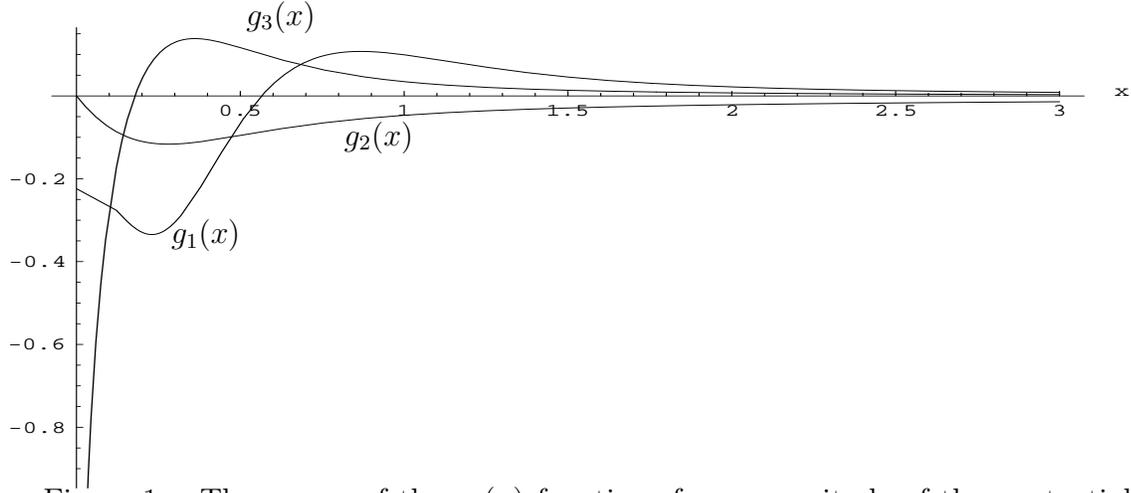,width=15cm,height=15cm}}
\put(2.7,5.6){$g_3(x)$}
\put(1.7,2.7){$g_1(x)$}
\put(4,4){$g_2(x)$}
\end{picture}
\caption{ The curves of the  $g_n(x)$ functions  for a magnitude of the 
pontential equal to $1$.} 
\end{figure}

For the complete quantum energy we still need the $E_f$ contribution. In the expression (26) for  $E_f$, after putting $s=0$, we integrate by parts and we obtain
\begin{equation}
E_f= \frac1\pi\sum^\infty_{l=0}(l+\frac12)\int^\infty_m \frac{k}{\sqrt{k^2-m^2}}\left(\ln f_\nu (ik) - \ln f^{as}_\nu (ik)\right) dk
\end{equation}
this quantity can't be further analytically simplified. Below we show (Fig. 2) a plot of $ R\cdot E_f$ as function of $R$ for $\alpha=1$.\\
\begin{figure}[ht]\unitlength1cm
\begin{picture}(7,7)
\put(0,-4){\epsfig{file=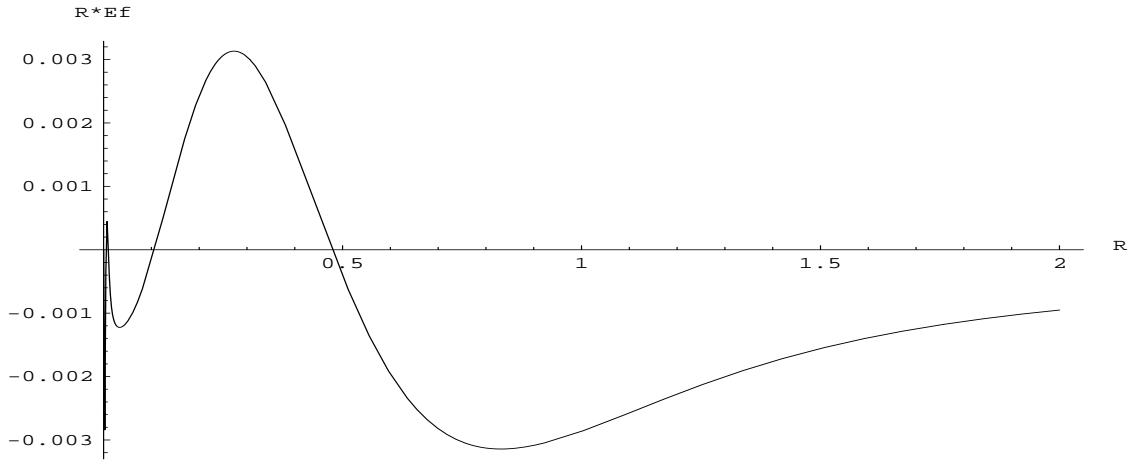,width=15cm,height=14cm}}
\end{picture}
\caption{ The curve of $R\cdot E_f(R)$ for a magnitude of the potential equal to 1. For $R=0$ the curve converges at $ R\cdot E_f\sim -3.9$.} 
\end{figure}

\vspace{0.5cm}

For the total ground state energy as a function of the radius of the shell we get the curves shown below (Fig. 3-5) for different values of the magnitude of the potential $\alpha$.
\vspace{1.5cm}

\begin{figure}[ht]\unitlength1cm
\begin{picture}(5,5)
\put(0,-4){\epsfig{file=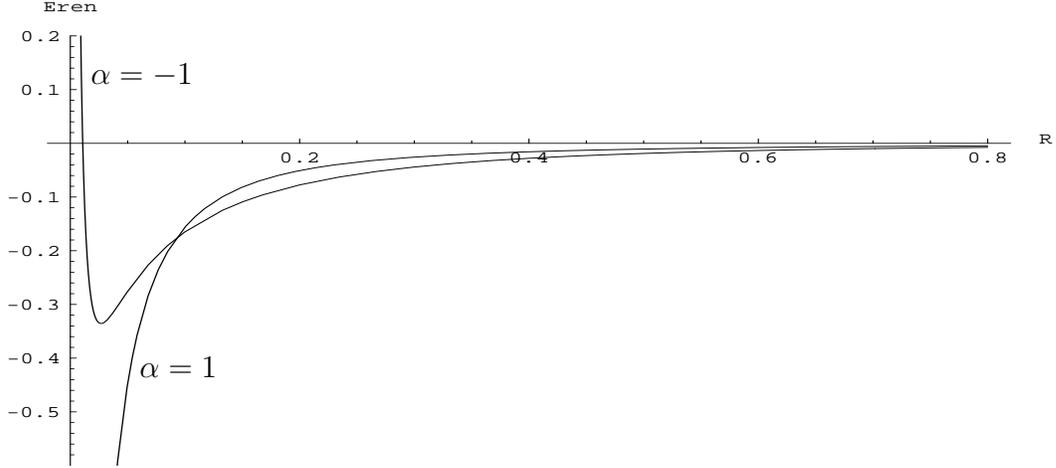,width=14cm,height=14cm}}
\put(1.8,1.1){$\alpha=1$}
\put(1.15,5){$\alpha=-1$}
\end{picture}
\caption{The renormalized vacuum energy $E^{ren}_\varphi (R)$ for positive and negative values of the potential.} 
\end{figure}

\begin{figure}[ht]\unitlength1cm
\begin{picture}(6,7)
\put(0,-4.1){\epsfig{file=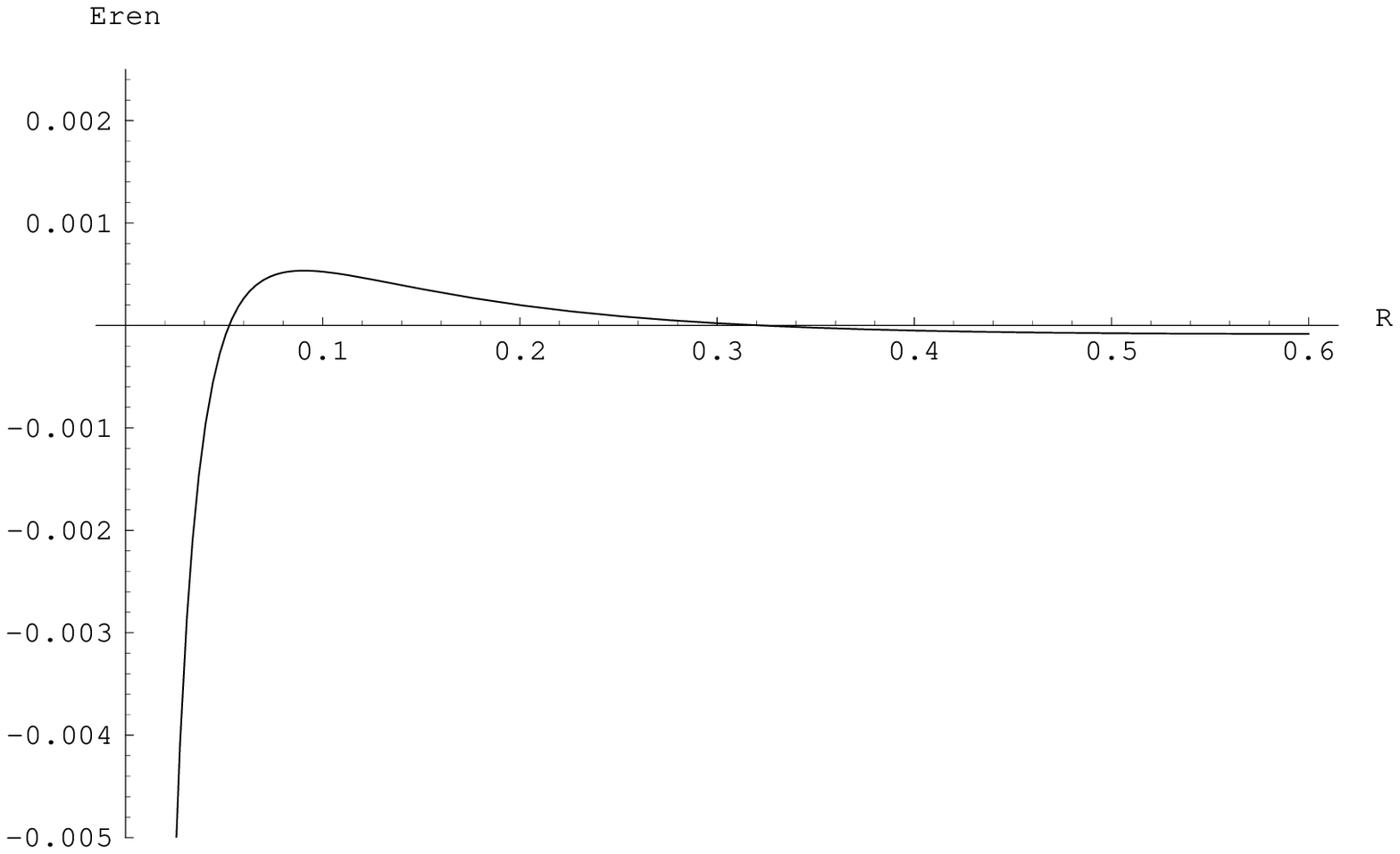,width=14cm,height=14cm}}
\end{picture}
\caption{The renormalized vacuum energy $E^{ren}_\varphi (R)$ for $\alpha$=0.3} 
\end{figure}

\begin{figure}[ht]\unitlength1cm
\begin{picture}(5,5)
\put(0,-4){\epsfig{file=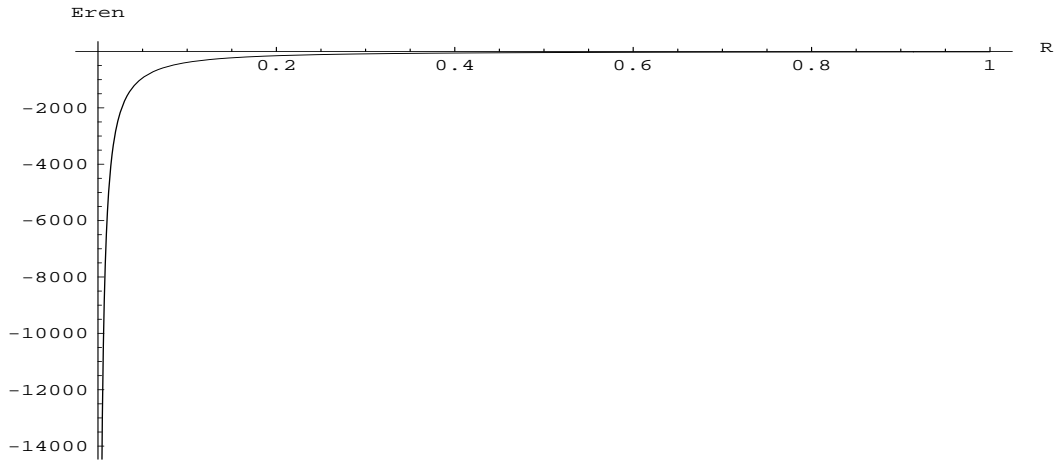,width=14cm,height=14cm}}
\end{picture}
\caption{The renormalized vacuum energy $E^{ren}_\varphi (R)$ for $\alpha$=10} 
\end{figure}

\section{Conclusion}
We have obtained an explicit expression for the renormalized ground state energy of a scalar massive field in the background of a semi-transparent shell. This expression is given by the sum of (49) and (51) and it depends only on the two parameters of the classical system, namely the radius and the magnitude of the potential of the spherical shell. 
The plots of $E^{ren}_\varphi $ as a function of the radius show that for repulsive potentials  the renormalized ground state energy  is positive only in some limited intervals of the radius and only when the potential is smaller than 1. For a potential larger than 1 the energy is always negative. This is the most striking conclusion of our work. For  very large values of $\alpha$ the shell should be no more transparent and  the problem should formally  become a Dirichlet boundary condition problem. One could check this in the equation (35) for the Jost function: here inserting a large $\alpha$ the addend $1$ becomes negligible, then we would just have the product of the two modified Bessel $I$ and $K$ functions;  such a Jost function is exactly the one obtained in \cite{3} for a perfectly reflecting spherical shell ( Dirichlet boundary conditions). In that paper the ground state energy is simply the sum of the energy inside and outside the shell. Then formally we should have:
\[
\lim_{\alpha \rightarrow +\infty}\ \  \ GSE^{semi-trans.} = \ \ \ GSE^{mirror}
\]

Now it is shown in \cite{3} that the ''mirror,, configuration has always  positive ground state energy for repulsive potentials. This fights with our plots which show an opposite sign. Furthermore the $A_2$ coefficient in paper \cite{3} is zero. In our work  $A_2$ remains a non zero coefficient also in the limit $\alpha \rightarrow +\infty$.\\ Nevertheless for flat parallel  boundaries the formal transition hypothized in the limit above is actually fulfilled as shown in paper \cite{9}.   At least at the present stage of the investigation it is difficult to interpret these results. We must stress however that a perfectly reflecting boundary is not a realistic object and that a semi-tansparent sphere is a better physical model.

\section{Acknowledgment}
 
I thank M. Bordag for advice.

\end{document}